\documentclass[12pt]{article}
\usepackage{epsfig, amssymb}
\usepackage{graphicx,epsfig}
\usepackage{color}
\begin{document}
\begin{titlepage}
\thispagestyle{empty}
\begin{flushright}
\end{flushright}

\bigskip

\begin{center}
\noindent{\Large \textbf
{Stochastic quantization and holographic Wilsonian renormalization group of free massive fermion}}\\
\vspace{2cm} \noindent{
Sung pil Moon\footnote{e-mail:mlnow@naver.com}}

\vspace{1cm}
  {\it
Department of Physics, Hanyang University, Seoul 133-791, Korea\\
 }
\end{center}

\vspace{0.3cm}
\begin{abstract}
We examine a suggested relation between stochastic quantization and the Holographic Wilsonian Renormalization Group in the massive fermion case on Euclidean AdS space. The original suggestion about the general relation between the two theories is posted in arXiv:1209.2242. In the previous researches, It is already verified that scalar fields, U(1) gauge fields, and massless fermions are consistent with the relation. In this paper, we examine the relation in the massive fermion case. Contrary to the other case, in the massive fermion case, the action needs particular boundary terms to satisfy boundary conditions. We finally confirm that the proposed suggestion is also valid in the massive fermion case.
\end{abstract}
\end{titlepage}

\newpage

\tableofcontents
\section{Introduction}
AdS/CFT correspondence has provided much understanding on various strongly coupled field theories by employing classical gravity theories defined on asymptotic AdS spacetime.  A partition function of a certain gravity theory on asymptotially AdS$_{d+1}$ spacetime is related to a partition function of a dual conformal field theory(CFT) defined on $d$-dimensional flat spacetime with a certain deformation. States and operators defined in the CFT correspond to normalizable and non-normalizable exitations in the gravitational theory respectively.\cite{4}, \cite{9}  For example, the field of scalar field theory in AdS$_{d+1}$ behaves as follows,
\begin{equation}
\phi(r)=\phi_0 r^{h_-} + \phi_1 r^{h_+},
\end{equation}
when the radial coordinate of AdS space `$r$' approaches its boundary i.e. $r\rightarrow0$, where $h_\pm=\frac{d}{4}\pm\frac{\sqrt{d^2+4m^2}}{4}$ and $m$ is the mass of the scalar field. The first term of the scalar field is non-normalizable, because $ r^{h_{-}}$ diverges near the $r\rightarrow0$ boundary.
Turnning on the non-normalizable mode plays a role of turnning on a dual operator on the boundary CFT and its coefficent $\phi_0$ corresponds to a source in the generating functional of the boundary CFT. On the contrary, the normalizable mode of the field is related to one state of the boundary CFT and the coefficient $\phi_1$ corresponds to a one-point function in the dual CFT.

The relation between the AdS theory and the boundary CFT is not only well defiend on the AdS boundary but also valid on any hypersurface defined on $r=\epsilon$.
In more detail, an AdS theory defined with a certain $r=\epsilon$ cut-off is related to a dual CFT defined on the $r=\epsilon$ hyper surface and having a certain cut-off energy scale.
In the AdS/CFT context, the AdS radial direction is related to the energy scale of the dual CFT, so the evolution of the AdS theory along the radial direction is connected to the Renormalization group flow of the dual CFT.

The relation between the cut-off parameters of the two theories can be understood by looking at the famous UV-IR relation.  The generating functional of the dual CFT gotten from AdS/CFT correspondence has many divergent elements when the AdS radial coordinate $r$ approaches to $0$ or $\infty$. The $r\rightarrow 0$ divergence is interpreted as the UV-divergence of the dual CFT which is proportional to a positive power of $\frac{1}{\epsilon}$ , and similarly, the $r\rightarrow \infty$ divergence is known to be related to the IR divergence of the dual CFT.

The correlation between the two theories can be understood from similarities between the Wilsonian Renormalization group and the holographic renormalization group.\cite{5}  For renormalizable QFT theory, Wilson separates fields into two parts depending on energy scale, $M=M(p^0<\Lambda)+M(\Lambda <p^0<\Lambda_0)$, and integrates out high energy fields $\Lambda_0>p^0>\Lambda$ as follows,

\begin{equation}
Z_{CFT}=\int DM_{p^0<\Lambda}DM_{\Lambda_0>p^0>\Lambda} e^{-S-\int \phi_0 \mathcal O }=\int DM_{p^0<\Lambda}e^{-S(\Lambda)-\int \tilde \phi \mathcal O (\Lambda)}
\end{equation}
where $M$ represent QFT boundary fields and $\Lambda_0$ is the UV cut-off and $\mathcal O$ represents a deformation defined on $p^0=\Lambda^0$. $\phi_0$, $\tilde \phi$ represent source terms at the scale $p^0=\Lambda^0$ and $p^0=\Lambda$ respectively. Because high energy quantities are intergrated, the quantities defined in the last equality have the information of high energy quantities. Thus the effective action $S(\Lambda)$ and the deformation $\mathcal O (\Lambda)$ at the scale $\Lambda$ are the function of the cut-off scale $\Lambda$.
The evolution of quantities of the QFT along the energy scale scale $\Lambda$ 
is called the Wilsonian Renormalization Group flow.
In the majority of cases, it is impossible to calculate the explicit solution of physical quantities of field theory but the Wilsonian Renormalization Group flow provides information of the cut-off $\Lambda$ in physical quantities, which helps in dimensional analysis. If one takes $\Lambda \rightarrow \infty$ limit, one could get information about the full path integral.


For gravity theory, likewise, one could separate fields along the radial cordinate $r$, $\phi=\phi(r>\epsilon)+\phi(\epsilon_0<r<\epsilon)$, and integrates out the short-length fields,

\begin{equation}
Z_{GRA}=\int D\phi_{\epsilon_0<r<\epsilon}D\phi_{r>\epsilon} e^{-S_{bulk}(\phi)}=\int D\phi_{r>\epsilon} e^{-S_{bulk}(\tilde\phi)-S_B(\tilde \phi)},
\end{equation}
where $\phi$ represent the bulk fields, $\tilde\phi(x)=\phi(r=\epsilon,x)$ , $S_B$ is the boundary action defined on $r=\epsilon$, and $S_{bulk}$ is the bulk action. As a result of the integration of the short-length fields, information of the integrated fields appears in the boundary action $S_B$ from the last equation. In the AdS/CFT context, the short-range part of the gravity theory of the first equation corresponds to the high energy part(UV-region) and classical limit of the total action $S_{tot}\equiv S_{bulk}+S_B$ corresponds to the Wilsonian effective action $S(\Lambda)$.\cite{6} As a result of the classical limit, the total action only contains effectively deformed part from the original bulk action, and the deformed part is related to the single trace operator $\mathcal O$ in the dual field theory by the AdS/CFT correspondence. More precisely, couplings of $\tilde \phi$ ,${\tilde \phi}^2$ , $\tilde \phi^3...$ in the boundary action $S_B$ correspond to expectation values of $\mathcal O$, $\mathcal O^2$ $\mathcal O^3...$ repectively.

Because the integrated degree of freedom contributes to the boundary action $S_B$, the boundary action gives boundary conditions for the bulk action  in the region $r>\epsilon$. By using variation principle, the boundary condition appears as a form of the relation between the conjugate momentum of the bulk field and the boundary action. Moreover, by using the fact that the total action $S_{tot}$ is invariant under the choice of arbitrary cut-off $\epsilon$, taking derivative with respect to $\epsilon$ gives evolution equation for $S_B$. The evollution equation is called the flow equation of the boundary action. From the definition of the conjugate momentum at the $r=\epsilon$ boundary, one can construct Hamiltonian from the Legendre transformation of the Lagrangian. As a result, the flow equation could be written as a form of the Hamilton-Jacobi equation \cite{1} ,
\begin{equation}
\label{Abcd}
\partial_\epsilon \psi_H(\phi,\epsilon)=-\int_{r=\epsilon}d^dx \mathcal H_{RG} \psi_H(\phi,r)
\end{equation}
where $H_{RG}$ is the Hamiltonian and $\psi_H\equiv e^{-S_B}$. Then the solution of the Hamilton-jacobi equation provides the forms of couplings on the $r=\epsilon$ hypersurface.

Stochastic quantization, on the other hand, is a different quantization method of fields apart from the other quantization methods of fields such as canonical or path integral quantization. \cite{8} 
 One unique feature of stochastic quantization is the existance of a new coordinate, a fictitious time $t$. Stochastic quantization is a technique quantizing a field theory by using relaxation process called Langevin equation, a type of the diffusion equation of a system coupled to a thermal reservoir. A stochastic system starts from an initial fictitious time $t=t_0$, where the system is in a non-equilibrium state, and as $t$ goes to infinity, the system settles down to equilibrium. The main idea of stochastic quantization is that the partition function of the equilibrium state has the same mathematical form of the partition function of Euclidean quantum field theory, so we can match the stochastic system in equilibrium with the Euclidean quantum system. In this equilibrium limit, statistical average values of physical quantities of the stochastic system correspond to Euclidean vacuum expectation values of the field theory.

In stochastic quantization, a new field appears, and it is called a noise field $\eta$. This field shows randomness, providing a property of the stochastic system, and the noise field mediates interaction between a large heat reservoir and the system. The best known example for such system is Brownian motion of a test particle. The particle's motion in this system is called Markovian process which is defined in the following situation : When a test particle collides with particles of the background fluids, the probability distribution of the particle's velocity after collision does not depends on its velocities in its earlier time, but only depends on the information just before the collision. In addition to the Markovian process, it is assumed that the probability distribution of the noise $\eta$ has the Gaussian form,

\begin{equation}
Z_{SQ}=\int D\eta \exp \left [-\frac{1}{4k}\int \eta^2(x,t)d^nxdt \right ]
\end{equation}
 and then the correlation functions between the noise fields are given by the delta-function form,
\begin{equation}
 \langle \eta(x,t)\eta(x',t') \rangle\propto \delta(x-x')\delta(t-t'),
\end{equation}
which means the noise field is affected by the information just before its current time.

 As mentioned above, the relaxation process of stochastic fields is called the Langevin equation. The form of the Langevin equation of the Braownian motion is

\begin{equation}
\label{Langevin equation}
\frac{\partial\phi(x,t)}{\partial t}=-k\frac{\partial S_E}{\partial \phi(x,t)}+\eta(x,t)
\end{equation}
where $\phi$ is a non-equilibrium stochastic field, $S_E$ is a non-equilibrium Euclidean action, and $k$ is a kernel used for convenience. 
The solution of the Langevin equation and the properties of the noise field are used to obtain expectation values of the stochastic fields.

 Contrary to the Langevin approach, one could represent the probability distribution with the stochastic field $\phi$ instead of the noise field $\eta$ by using the Langevin equation. In fact, this process is a path integral with the stochastic fields, and the action of the path integral is called the Fokker-Plank action $S_{FP}$,
\begin{equation}
\label{SQ action}
Z_{SQ}=\int [D\phi(0)]e^{-\frac{S_{E}[\phi(0)]}{2}}\int [D\phi]e^{-S_{FP}}
\end{equation}
 where the Fokker-Plank action is written as

\begin{equation}
\label{Fokker-plank equation}
S_{FP}=\int_0^T dt \int d^dx[\frac{1}{4k}\dot \phi+\frac{1}{4}k(\frac{\delta S_{cl}}{\delta \phi})^2-\frac{1}{2}k\frac{\delta^2 S_{cl}}{\delta \phi^2}],
\end{equation}
\begin{equation}
\label{Fokker-plank equation}
[D\phi]=\prod_{0<t<T} [D\phi(t)].
\end{equation}
The fictitious time starts from $t=0$ where system is in a non-equilibrium state and reaches to $t=T$ where system settles down in the equilibrium state.


In the last 10 years similarities between the two theories, Holographic Wilsonian Renormalization Group(HWRG) and Stochastic quantization, have been observed.  Especially the author of \cite{7} links specifically between quantities of the two theories. The author sets the initial time to $t=-T$ and sets the time when the system settles into equilibrium to $t=0$ in stochastic quantization. Moreover the author newly defines a HWRG action as follows,
\begin{equation}
\label{newly defined hol action}
Z'=\int [D \phi_0]e^{-I_d[\phi_0]} Z_{GRA}
\end{equation}
where $I_d$ is the classical limit of the total action and it is used for the Boltzmann weight.
Now the new partition function has boundary fields and bulk fields as integral elements. This feature can be captured by stochastic partition function configuration (\ref{SQ action}). The author shows that if one equate $Z'$ and $Z_{SQ}$, there is one to one correspondence between $S_{FP}$, $S_{E}$, $t$ in the stochastic quantization and $S_{tot}$, $2I_d$, $r$ in the HWRG. 
 Moreover the Fokker-Plank equation of motion has its Hamiltonian formalism,

\begin{equation}
\label{int_wyel fermion}
\partial_t \psi_S(\phi,t)=-\int d^dx H_{FP} \psi_S(\phi,t)
\end{equation}
where $\psi_S=P(\phi,t)e^{\frac{S_c[\phi(t)]}{2}}$ and$P(\phi,t)$ is the probability distribution of the stochatic fields, which can be matched by Hamiltonian equation of motion of AdS/CFT (\ref{Abcd}). 

From the previous discovery of the similar mathematical structure between the two theories, another specific relationship between the two theories which our paper follows is suggested \cite{1}. Here are some of the details. (1) The classical action in stochastic quantization and the on-shell action in AdS/CFT are related with $S_c=-2I_{os}$. (2) Equate the stochastic fictitious time $t$ with the radial coordinate $r$ of AdS/CFT.  (3) Hamiltonians of the two theories are same. The results of the suggestion are summarized in two following parts.

Firstly, the AdS boundary action is related to the stochastic Fokker-plank lagrangian,

\begin{equation}
\label{int_wyel fermion}
 S_B=\int ^{t}dt'\int d^dx L_{FP}[\phi(t,x)]
\end{equation}
where $L_{FP}$ is the Fokker-Plank Lagrangian.

 Secondly, the two point correlations of AdS/CFT and stochastic quantization are related as follows,

\begin{equation}
\label{int_wyel fermion}
<\phi_q(t)\phi_{q'}(t)>^{-1}_{H}=<\phi_q(t)\phi_{q'}(t)>^{-1}_{S}-\frac{1}{2}\frac{\delta^2S_c}{\delta \phi_q(t) \delta_{q'}(t)}
\end{equation}
where $S_c$ is the Euclidean classical action of the stochastic quantization.
 The relations between the two theories are already examined with several examples, for example, scalar fields, vector bosons, and massless fermions. The results are consistent with the suggestion\cite{2,3}.

In this paper, we extend the massless fermion case to massive fermions, and show that the the relation between the two theories are satisfied, so the massive fermion case is also consistent with the suggestion. Overall procedures are similar to the previous massless case. Thus we use similar notation with that of the previous massless fermion case. All calculations are performed in Euclidean signature.

In previous massless fermion case, the fermion action needs additional boundary terms to impose the Dirichlet boundary condition or the Neumann boundary condition on the boundary consistently. Likewise,  we add an additional boundary terms which give the Dirichlet boundary condition and the Neumann boundary condition consistently into the fermion action,
\begin{equation}
\label{221}
S_{add}=-\left.\left.i\int_{r=\epsilon} d^d k  \right[\bar \chi_+(k) \chi_-(-k)-i\bar \chi_+(k) \frac{m}{r}\frac{k^{\hat\mu} \gamma_{\mu}}{|k|^2}\chi_+(-k) \right],
\end{equation}
where the first term is also used in the previous massless case. However the first term gives only the Dirichlet boundary in the massive case, so we add the second term to impose the Neumann boundary condition. As you know, the additional boundary terms do not affect the overall equation of motion of fermions, so these terms can solve the boundary condition problem smoothly.

In Section 2, we obtain the explicit solutions of the fermionic field and the double trace deformation in the AdS/CFT case with our notation. The calculation is already well performed in \cite{11}. The solutions are similar to the solution of the massless case, but several non-trivial terms are included as well. For instance the double trace deformation is 
\begin{eqnarray}
\label{int_double massive}
\label{Doubkl-tra-couplking}
D_{\alpha\beta}(\epsilon,k)&=&
\frac{E_\alpha(k)}{|k|^2}\delta_{\alpha\beta}\left(|k|\frac{K_{m+\frac{1}{2}}(|k|\epsilon)-\Delta_{\alpha,+}I_{m+\frac{1}{2}}(|k|\epsilon)}{K_{m-\frac{1}{2}}(|k|\epsilon)
 +\Delta_{\alpha,+}I_{m-\frac{1}{2}}(|k|\epsilon)}-\frac{m}{r} \right),
\end{eqnarray}
where $K$ and $I$ are modified bessel functions. The first term in the parenthesis is similar to the previous solution, but the second term is new to the previous soltuion which goes to zero when the mass is zero.
 In section 3, the stochastic calculation begins with the Euclidean action $E_c$ obtained from the on-shell action in Section 2. We first follow the Langevin approach, and match the result with the previous AdS/CFT case, and identify the suggested relation,
\begin{equation}
\langle \chi_\alpha(k,t) \bar \chi_\beta(k,t) \rangle_H^{-1} = \langle \chi_\alpha(k,t) \bar \chi_\beta(k,t) \rangle_S^{-1}-\frac{1}{2} \frac{\overrightarrow\delta}{\delta \bar \chi_\alpha(-k,t)} 
S_c \frac{\overleftarrow\delta}{\delta \chi_\beta(-k,t)}.
\end{equation}
Lastly we calculate the Fokker-Plank action, and ensure that the Fokker-Plank Hamiltonian is exactly the same with the AdS/CFT Hamiltonian. Double trace deformation gotten from Fokker-Plank approach also gives the same form with the double trace deformation of AdS/CFT in (\ref{int_double massive}).

\section{Holographic Wilsonian Renormalization Group of massive fermions}
\label{section2}
In this section we review the Holographic Wilsonian Renormalization Group of massive fermions in Euclidean $AdS_{d+1}$ space. Calculations and discussions about the HWRG of bulk fermions are already performed in \cite{10,11} and we use similar notations with them. A discussion about the boundary effective action for fermions is also well performed in \cite{10,11}
\subsection{Setup for explicit calculation}

	The fermionic action in Euclidean AdS space is written as a bulk action $S_{bulk}$ together with a boundary action $S_B$,
\begin{equation}
\label{HWRG action}
S_{tot}=S_{bulk}+S_B(\psi ,\bar\psi).
\end{equation}
where the AdS boundary is located on the $r=\epsilon$ hypersurface and the boundary action is a functional of the fermionic fields $\psi(\epsilon)$ and $\bar\psi(\epsilon)$ on the boundary.
     The metric of the Euclidean $AdS_{d+1}$ is 

\begin{equation}
\label{Euclidean metric}
ds^2=\frac{1}{r^2}\left( dr^2 + \sum_{\mu,\nu=1}^d\delta_{\mu\nu}dx^\mu dx^\nu \right),
\end{equation}
where `$r$' represents AdS radial coordinate, and the Greek indices on the coordinates $x$ represent the AdS boundary indicies.
We start with the familiar fermion bulk action in AdS space,
\begin{equation}
\label{fermion action}
S_{bulk}=-i\int_{r>\epsilon} drd^dx \sqrt g(\bar\psi\Gamma^M\nabla_M\psi-m\bar\psi\psi).
\end{equation}
where $g$ is the determinant of the metric of the Euclidean AdS space, $m$ is the mass of the field $\psi$, and the fermionic field $\bar\psi$ is related to $\psi$ with the relation $\bar\psi=\psi^\dagger\Gamma^{\hat 0}$. M represents the full Euclidean AdS coordinates, and the covariant derivative $\nabla_M$ is defined by

\begin{equation}
\label{Covariant derivative}
\nabla_M=\partial_M+\frac{1}{4}\omega^{\hat A \hat B}_M\Gamma_{\hat A\hat B},
\end{equation}
where $\omega^{\hat A\hat B}_M$ is a spin connection. $\Gamma_{\hat A\hat B}$ is a generator which rotates spinors on the tagent space

\begin{equation}
\label{non-abelian gauge transformation}
\psi\rightarrow e^{i\epsilon^{\hat A\hat B} \Gamma_{\hat A\hat B}}\psi ,  {\rm \ \ and  \ \ }\bar\psi\rightarrow e^{i\epsilon^{\hat A\hat B} \Gamma_{\hat A\hat B}}\bar\psi .
\end{equation}
where $\hat A$ and $\hat B$ represent the tangent space indices, $\epsilon^{\hat A\hat B}$ is the rotation angle. The generator $\Gamma_{\hat A\hat B}$  is defined as $\Gamma_{\hat A\hat B}\equiv\Gamma_{\hat A}\Gamma_{\hat B}-\Gamma_{\hat B}\Gamma_{\hat A}$ where $\Gamma_{\hat A}$ are gamma matrices of the tangent space. For the case where the boundary space dimension $d$ is odd, thus the bulk space dimension is even, $\Gamma^M$ is defined by
\begin{eqnarray}
\label{gamma matrices}
\Gamma^{\hat r}=\left(\begin{array}{cc} \mathbb I & 0 \\ 0 & -\mathbb I\end{array}\right) {\rm,\ \ and \ \ }
\Gamma^{\hat \mu}=\left(\begin{array}{cc} 0 & \gamma^{\hat \mu} \\ \gamma^{\hat \mu} & 0 \end{array}\right).
\end{eqnarray}
where $\gamma^{\hat\mu}$ are the gamma metrices of the boundary space-time, and $\mathbb I$ is the identity matrix. The size of $\gamma^{\hat\mu}$ and $\mathbb I$ is $2^{\frac{d-1}{2}}\times 2^{\frac{d-1}{2}}$. For the case where the boundary space dimension $d$ is even, thus the bulk space dimension is odd, $\Gamma^{\hat r}$ is defined by $\Gamma^{\hat r}\equiv\gamma^{\hat 0}\gamma^{\hat 1}...\gamma^{\hat d-1}$ and $\Gamma^{\hat \mu}$ is defined by $\Gamma^\mu\equiv\gamma^{\hat \mu}$ where the size of gamma matrices is $2^{\frac{d}{2}}\times 2^{\frac{d}{2}}$.

 We use the same notation with \cite{10}  for fermionic fields, $\Phi=(gg^{rr})^{1/4} \psi$ and $\bar\Phi=(gg^{rr})^{1/4} \bar\psi$  for convenience. With the newly defined fermionic fields and after trivial calculation, the bulk action is written as

\begin{equation}
\label{redefined action}
S_{bulk}=-i\int dr d^d x \left( \bar\Phi \Gamma^{\hat r} \partial_r \Phi + \bar\Phi \Gamma^{\hat \mu} \partial_{\mu} \Phi  - \sqrt{g_{rr}}m\bar\Phi\Phi\right),
\end{equation}
where the hat indices represent the tangent space indices.
We could decompose the fermionic fields  into $\chi$ and $\bar\chi$ fields, with the right-handed projection operator $P_+=(1+\Gamma^{\hat r})/2$ and the left-handed projection operator $P_-=(1-\Gamma^{\hat r})/2$. So the fermionic fields are classified into four kinds of fields
\begin{equation}
\label{weyl fermion}
\chi_\pm=\frac{1\pm\Gamma^{\hat r}}{2}\Phi {\rm , \ \ and \ \ }\bar\chi_\pm=\frac{1\mp\Gamma^{\hat r}}{2}\bar\Phi.
\end{equation}
If the boundary space dimension $d$ is odd, $\chi$ and $\bar \chi$ fields are Dirac fermions on the boundary, but are Weyl fermions in the bulk. So the fermionic field $\Phi$ can be written as decomposed form,

\begin{eqnarray}
\label{field}
\Phi=\left(\begin{array}{cc} \chi_+ \\ \chi_-\end{array}\right).
\end{eqnarray}
On the other hand, if the boundary space dimension $d$ is even, $\chi$ and $\bar \chi$ are Weyl fermions on the boundary, but are Dirac fermions in the bulk.

\subsection{Equations of motion and the solutions of the fermionic fields}

We put the $\chi$ and $\bar \chi$ fields into the bulk action (\ref{redefined action}),
\begin{eqnarray}
\label{chi action}
S_{bulk}&=&\left.-i\int_{r>\epsilon} dr d^dk \right[-\bar \chi_{+}(k)\partial_r \chi_-(-k) + \bar\chi_-(k)\partial_r \chi_{+}(-k)  \\
 \nonumber
&-&ik_{\mu} \bar\chi_+(k)\gamma^{\hat \mu}\chi_+(-k)- ik_{\mu}\left. \bar\chi_-(k)\gamma^{\hat \mu}\chi_-(-k)-\frac {m}{r}\bar\chi_+(k)\chi_-(-k)-\frac {m}{r}\bar\chi_-(k)\chi_+(-k)\right].
\end{eqnarray}

By applying the principle of least action to the bulk action, we can get the equations of motion for $\chi$ and $\bar\chi$ fields,

\begin{eqnarray}
\label{24}
0&=&\partial_r \chi_+(-k)-ik_\mu\gamma^{\hat \mu}\chi_-(-k)-\frac{m}{r}\chi_+(-k), \\
\label{25}
0&=&-\partial_r \chi_-(-k)-ik_\mu\gamma^{\hat \mu}\chi_+(-k)-\frac{m}{r}\chi_-(-k), \\
\label{26}
0&=&-\partial_r \bar\chi_-(k)-ik_\mu\bar\chi_+(k)\gamma^{\hat \mu}-\frac{m}{r}\bar\chi_-(k), \\
\label{27}
0&=&\partial_r \bar\chi_+(k)-ik_\mu\bar\chi_-(k)\gamma^{\hat \mu}-\frac{m}{r}\bar\chi_+(k).
\end{eqnarray}
There are four kinds of the equations, and one would think they are independent each other. However the fermionic fields have the relation $\bar \chi_\pm=\chi_\pm^\dagger\gamma^{\hat 0}$, so there are two independent equations. If we combine the eqauation (\ref{24}) and (\ref{25}), we can get the equation for $\chi_+$,

\begin{equation}
\label{explicit equation}
r^{-m}\partial_r(r^{2m}\partial_r(r^{-m}\chi_{\alpha,+}(k)))-{|k|}^{2}\chi_{\alpha,+}(k)=0,
\end{equation}
where $\alpha$ are spin indices and $|k|^2\equiv k^\mu k_\mu$.
The solution of the $\chi_+$ equation has a form of linear combination of modified bessel functions $K_{m-\frac{1}{2}}(|k|r)$ and $I_{m-\frac{1}{2}}(|k|r)$. The general soultion of the $\chi_+$ equation is given as

\begin{equation}
\label{explicit solution for chi}
\chi_{\alpha,+}(k,r)=\chi^{(0)}_{\alpha,+}(k)r^{\frac{1}{2}}K_{m-\frac{1}{2}}(|k|r)+\chi^{(1)}_{\alpha,+}(k)r^{\frac{1}{2}}I_{m-\frac{1}{2}}(|k|r),
\end{equation}
where  $\chi^{(0)}_{\alpha,+}$ and  $\chi^{(1)}_{\alpha,+}$ are coefficient spinors which are determined by boundary condition.

We could diagonalize $k_\mu\gamma^\mu$ because we use the Euclidean metric and therefore $k_\mu\gamma^\mu$ is hermitian matrix as follows

\begin{equation}
\label{diagonalization}
[k_\mu\gamma^\mu]_{\alpha\beta}=E_{\alpha}(k)\delta_{\alpha\beta},
\end{equation}
where $E_\alpha(k)$ is the eigenvalue of a diagonalized eigenvector $|\alpha\rangle$ where

\begin{equation}
\label{eigen vector}
|\alpha\rangle=
\left(\begin{array}{cccc}0  \\ 0... \\1\\... 0 \\ 0  \end{array}\right) \leftarrow {\rm\ \alpha th\ row}.
\end{equation}
Then we can write the solution of the $\chi_+$ equation as an eigenvector form,

\begin{eqnarray}
\label{vector field}
\chi_{\alpha,+}(k,r) &=&N_{\alpha,+}r^{\frac{1}{2}}
\left(\begin{array}{cccc}0  \\ 0... \\{K_{m-\frac{1}{2}}(|k|r)+\Delta_{\alpha,+}(k)I_{m-\frac{1}{2}}(|k|r) } \\... 0 \\ 0  \end{array}\right) \leftarrow {\rm\ \alpha th\ row}, \\ \nonumber
&\equiv&N_{\alpha,+}r^{\frac{1}{2}} [K_{m-\frac{1}{2}}(|k|r)+\Delta_{\alpha,+}(k)I_{m-\frac{1}{2}}(|k|r) ]  |\alpha\rangle,
\end{eqnarray} 
where $N_{\alpha,+}$ and $\Delta_{\alpha,+}$ are coefficient functions of $k$, which are also determined by boundary condition.

Like the $\chi_+$ field, If we combine the quation (\ref{24}) and (\ref{25}) with respect to the $\chi_-$ field, we can get the equation for the $\chi_-$ field and the solution, and the result is similar to the case for the $\chi_+$ field,
\begin{equation}
\label{214}
\chi_{\alpha,-}(k,r)=\chi^{(0)}_{\alpha,-}(k)r^{\frac{1}{2}}K_{m+\frac{1}{2}}(|k|r)+\chi^{(1)}_{\alpha,-}(k)r^{\frac{1}{2}}I_{m+\frac{1}{2}}(|k|r),
\end{equation}
where $\chi_{\alpha,-}^{(0)}$ and $\chi_{\alpha,-}^{(1)}$ are coefficient spinors. The soltuion is also written as a vector form as follows
\begin{equation}
{\rm  \ \ }
\label{215}
\chi_{\alpha,-}(k,r) =N_{\alpha,-}r^{\frac{1}{2}}({K_{m+\frac{1}{2}}(|k|r)+\Delta_{\alpha,-}(k)I_{m+\frac{1}{2}}(|k|r) })|\alpha\rangle,
\end{equation}
where $N_{\alpha,-}$ and $\Delta_{\alpha,-}$ are coefficient functions of $k$.

We can get relations between the coefficients of the $\chi_+$ and $\chi_-$ fields by putting the solutions into the equaions of motion (\ref{24}) and (\ref{25}),
 \begin{equation}
\label{216}
 \chi^{(0)}_{\alpha,+}(k)=\frac{ik_\nu \gamma^{\hat\nu}}{|k|}\chi^{(0)}_{\alpha,-}(k) {\rm\ \ and \ \ } \chi^{(1)}_{\alpha,+}(k)=-\frac{ik_\nu \gamma^{\hat\nu}}{|k|}\chi^{(1)}_{\alpha,-}(k) .
 \end{equation}

Similarly we could get the solutions for the $\bar\chi_{\pm}$ fields by combining the equations of motion (\ref{26}) and (\ref{27}),
\begin{eqnarray}
\label{217}
\bar\chi_{\alpha,+}(k,r)&=&\bar\chi^{(0)}_{\alpha,+}(k)r^{\frac{1}{2}}K_{m-\frac{1}{2}}(|k|r)+\bar\chi^{(1)}_{\alpha,+}r^{\frac{1}{2}}(k)I_{m-\frac{1}{2}} (|k|r), \\ 
\label{218}
{\rm and \ \ }
\bar\chi_{\alpha,-}(k,r)&=&\bar\chi^{(0)}_{\alpha,-}(k)r^{\frac{1}{2}}K_{m+\frac{1}{2}}(|k|r)+\bar\chi^{(1)}_{\alpha,-}(k)r^{\frac{1}{2}}I_{m+\frac{1}{2}}(|k|r),
\end{eqnarray}
where $\bar \chi_{\alpha,\pm}^{(0)}$ and $\bar \chi_{\alpha,\pm}^{(1)}$ are coefficient spinors determined by boundary condition.
The coefficient spinors of the $\bar\chi_\pm$ fields are related by the equations of motion (\ref{26}) and (\ref{27}),
\begin{equation}
\label{219}
\bar\chi^{(0)}_{\alpha,+}(k)=-\frac{ik_\nu }{|k|} \bar\chi^{(0)}_{\alpha,-}(k)\gamma^{\hat\nu}
{\rm\ \ and \ \ } \bar\chi^{(1)}_{\alpha,+}(k)=\frac{ik_\nu }{|k|}\bar\chi^{(1)}_{\alpha,-}(k) \gamma^{\hat\nu}.
\end{equation}

\subsection{Additional boundary terms and boundary conditions}

There are total derivative boundary terms which are left from the principle of least action principle given by
\begin{equation}
\label{boundary variation term2}
\delta S_{bulk}=i\left.\left.\int_{r=\epsilon} d^dk \right[ \bar \chi_+(k)\delta \chi_-(-k) - \bar \chi_-(k)\delta \chi_+(-k)\right].
\end{equation}
To remove the total derivative term, we should impose the Dirichlet boundary condition to the $\chi_+$ field, and also to the $\chi_-$ field. But two kinds of fields are related by the equations of motion, imposing the Dirichlet boundary condition to both side is inconsistent. Because additional boundary terms don't affect the equations of motion, we could add the following boundary terms to the original bulk action

\begin{equation}
\label{221}
S_b=-\left.\left.i\int_{r=\epsilon} d^d k  \right[\bar \chi_+(k) \chi_-(-k)-i\bar \chi_+(k) \frac{m}{r}\frac{k^{\hat\mu} \gamma_{\mu}}{|k|^2}\chi_+(-k) \right].
\end{equation}

In fact, the first term is sufficient If we just want to impose the Dirichelt boundary condition only. However we can also impose the Neumann boundary condition with the second term. Together with the additional boundary terms, the variation of the total action becomes

\begin{eqnarray}
\label{222}
\delta(\hat S)&=&\left.-i\int_{r=\epsilon} d^dk \right[ \delta\bar \chi_+(k) \chi_-(-k) + \bar \chi_-(k)\delta \chi_+(-k)-i\delta \bar \chi_+(k) \frac{m}{r}\frac{k^{\hat\mu} \gamma_{\mu}}{|k|^2}\chi_+(-k) \\ \nonumber
&-&\left. i\bar\chi_+(k) \frac{m}{r}\frac{k^{\hat\mu} \gamma_{\mu}}{|k|^2}\delta\chi_+(-k) \right],
\end{eqnarray}
and the boundary terms vanish by imposing the Dirichlet boundary condition or the Neumann boundary condtion on $\chi_+$ fields. The fermionic bulk action is newly defined as the sum of the original bulk action and the aditional boundary term

\begin{equation}
\label{223}
 \hat S=S_{bulk}+S_b.
\end{equation}

By using the equations of motion, we can substitute the $\chi_+$ and $\bar\chi_+$ fields for the $\chi_-$ and $\bar\chi_-$ fields in the total action. In other words, we can write the total action in terms of the $\chi_+$ and the $\bar\chi_+$ fields only,
\begin{equation}
\label{224}
S_{tot}=\hat S+S_B(\chi_+,\bar\chi_+),
\end{equation}
where the new bulk action $\hat S$  is written with with $\chi_+$ and $\bar \chi_+$ fields
\begin{eqnarray}
\label{225}
\hat S&=&-\int drd^dk\left[ \partial_r \bar\chi_+(k,r)\frac{k_{\hat \mu}\gamma^{\hat \mu}}{|k|^2}\partial_r\chi_+(-k,r)+\bar\chi_+(k,r)k_{\hat \mu}\gamma^{\hat \mu}\chi_+(-k,r)\right. \\ \nonumber
&+&\left. \frac{m(m-1)}{r^2}\bar \chi_+(k,r)\frac{k_{\hat \mu}\gamma^{\hat \mu}}{|k|^2}\chi_+(-k,r)\right].
\end{eqnarray}

\subsection{The Flow equation of fermion and the two-point correlation}
We get the conjugate momentums of $\chi_+$ and $\bar\chi_+$ fields from the variation of the total action,
\begin{eqnarray}
\label{227}
\Pi_+(k)&\equiv& \frac{\overrightarrow \delta \hat S}{\delta \partial_r \bar\chi_+(-k)}=-\frac{k_{\mu}\gamma^{\hat\mu}}{|k|^2}\partial\chi_+(k)= \frac{\overrightarrow \delta S_B}{\delta \bar\chi_+(-k)}, \\ \nonumber
\bar \Pi_+(k)
&\equiv& \frac{\overleftarrow \delta \hat S}{\delta \partial_r \chi_+(-k)}=-\frac{k_{\mu}\gamma^{\hat\mu}}{|k|^2}\partial\bar\chi_+(k)= \frac{\overleftarrow \delta S_B}{\delta \chi_+(-k)},
\end{eqnarray}
where the third equations of each array come after varying the total action and applying the principle of least action.
We get the flow equation of the boundary action by derivating the equation (\ref{224}) with respect to the radial cut-off $\epsilon$,
\begin{eqnarray}
\label{226}
\partial_\epsilon S_B&=& \int_{r=\epsilon} d^dk
\left[\left( \frac{\overleftarrow\delta S_B}{\delta \chi_+(-k)}\right)k_\mu\gamma^{\hat\mu} \left(\frac{\overrightarrow\delta S_B}{\delta \bar\chi_+(k)}\right) \right.\\ \nonumber
&-&\left. \bar\chi_+(k)k_\mu\left(1+\frac{m(m-1)}{r^2|k|^2}\right) \gamma^{\hat \mu} \chi_+(-k)\right],
\end{eqnarray}
To solve the flow equation, we use a trial functional of $S_B$ given by
\begin{equation}
\label{228}
S_B=\Lambda(\epsilon)+\left.\left.\int d^d k \right[ \bar J(k,\epsilon)\chi_+(-k) + \bar\chi_+(k) J(-k,\epsilon)+\bar \chi_+(k)D(k,\epsilon)\chi_+(-k)\right],
\end{equation}
where $\Lambda$ is constant, and $J$ and $\bar J$ and $D$ are functions of $k$.
As a results of putting the trial functional into the flow equation (\ref{226}), we get four equations about the coefficients,
\begin{eqnarray}
\label{229,230,231,232}
\partial_\epsilon \Lambda(\epsilon)&=& \int d^dk\bar J(k,\epsilon)\gamma^{\hat\mu}k_{\mu}J(-k,\epsilon), \\
\partial_\epsilon \bar J(k,\epsilon)&=& \bar J(k,\epsilon)\gamma^{\hat\mu}k_{\mu} D(k,\epsilon), \\
\partial_\epsilon J(-k,\epsilon)&=& D(k,\epsilon)\gamma^{\hat\mu}k_{\mu} J(-k,\epsilon), \\
{\rm and \ \ \ }
\partial_\epsilon D(k,\epsilon)&=& -k_{\mu}\gamma^{\hat\mu}\left(1+\frac{m(m-1)}{r^2|k|^2}\right)+D(k,\epsilon) k_\mu \gamma^{\hat\mu} D(k,\epsilon).
\end{eqnarray}
In contrast with the other equations, the last equation is written only with the coefficient of $\chi^2$, so we solve the last equation first. The form of the solution is easily inferred from \cite{6}, 
\begin{eqnarray}
\label{233}
D(k,\epsilon)&=&-\sum_{\alpha,\beta}\frac{[k_{\nu}\gamma^{\hat\nu}]_{\alpha\beta}}{|k|^2}\partial_\epsilon \chi_{\beta,+}(k,\epsilon) \chi^{-1}_{\alpha,+}(k,\epsilon), \\ \nonumber
&=&-\sum_\alpha \frac{E_\alpha(k)}{|k|^2}\partial_\epsilon \chi_{\alpha,+}(k,\epsilon) \chi^{-1}_{\alpha,+}(k,\epsilon),
\end{eqnarray}
\begin{equation}
{\rm where \ \ }
\label{234}
\chi^{-1}_{\alpha,+}(k,\epsilon)=\frac{1}{N_{\alpha,+}r^{\frac{1}{2}} (K_{m-\frac{1}{2}}(|k|\epsilon)+\Delta_{\alpha,+}(k)I_{m-\frac{1}{2}}(|k|\epsilon) )}\langle \alpha|.
\end{equation}
We get the explicit form of the $\chi^2$ coefficient by putting the solution of the $\chi_+$ fields into the $\chi^2$ coefficient,
\begin{eqnarray}
\label{236}
\label{Doubkl-tra-couplking}
D_{\alpha\beta}(\epsilon,k)&=&
\frac{E_\alpha(k)}{|k|^2}\delta_{\alpha\beta}\left(|k|\frac{K_{m+\frac{1}{2}}(|k|\epsilon)-\Delta_{\alpha,+}I_{m+\frac{1}{2}}(|k|\epsilon)}{K_{m-\frac{1}{2}}(|k|\epsilon)
 +\Delta_{\alpha,+}I_{m-\frac{1}{2}}(|k|\epsilon)}-\frac{m}{r} \right).
\end{eqnarray}
The coefficient of $\chi^2$  is related with the two point correlation function of the dual CFT from the AdS/CFT correspondence as follows,

\begin{equation}
\label{The relation between TPCF and DTO}
\langle \chi_{\alpha,+}(k,t) \bar \chi_{\beta,+}(k,t) \rangle_H^{-1}=D_{\alpha \beta}.
\end{equation}

\section{Stochastic quantization of massive fermions}
\label{Section 3}

\paragraph{Classical action}
The on shell action $I_{os}$ can be gotten from putting the equations of motion into the fermion action.
Because the equations of motion make the original bulk action disappear, the rest terms are just the additional boundary terms which are given as
\begin{equation}
\label{337}
I_{os}=S_b=-\left.\left.i\int_{r=\epsilon} d^d k  \right[\bar \chi_+(k) \chi_-(-k)-i\bar \chi_+(k) \frac{m}{r}\frac{k^{\hat\mu} \gamma_{\mu}}{|k|^2}\chi_+(-k) \right],
\end{equation}

Because the first kind modified bessel function $I_{m-\frac{1}{2}}$ diverges at $\epsilon=0$, the forms of the $\chi_+$ fields are made up with the second kind modified bessel function,
\begin{equation}
\label{338}
\chi_{+}({k,r})=\chi^{(0)}_{+}(k)K_{m-\frac{1}{2}}(|k|r),
\end{equation}
 
\begin{equation}
{\rm and \ \ }
\label{339}
\bar\chi_{+}({k,r})=\bar\chi^{(0)}_{+}(k)K_{m-\frac{1}{2}}(|k|r).
\end{equation}

Then the on-shell action can be written with the $\chi_+$ and $\bar\chi_+$ fields as follows,

\begin{equation}
\label{340}
 I_{os}=-\int d^d k\bar\chi_+(k,\epsilon)\left(\frac{k_\mu \gamma^{\hat \mu}}{|k|^2} \right)\chi_+(-k,\epsilon)\partial_{\epsilon}\log\left({\epsilon}^{\frac{1}{2}}K_{m-\frac{1}{2}}(|k|\epsilon)\right).
\end{equation}

The start point of stochastic quanziation is the suggestion $S_c=-2I_{os}$. So the classical action of the Stochastic quantization $S_c$ is given by

\begin{equation}
\label{341}
\label{the-classical-ActioN}
 S_c=2\int d^d k\bar\chi_+(k,\epsilon)\left(\frac{k_\mu \gamma^{\hat \mu}}{|k|^2} \right)\chi_+(-k,\epsilon)\partial_{\epsilon}\log\left({\epsilon}^{\frac{1}{2}}K_{m-\frac{1}{2}}(|k|\epsilon)\right).
\end{equation}

\subsection{Langevin approach}

The general forms of the langevin equation for fermion fields are

\begin{equation}
\label{342}
\frac{\partial \chi_+(k,t)}{\partial t}=-V(k)\frac{1}{2}\frac{\overrightarrow{\delta}S_c}{\delta \bar\chi_+(k,t)}+\eta(k,t),
\end{equation}

\begin{equation}
\label{343}
\frac{\partial \bar\chi_+(k,t)}{\partial t}=-V(k)\frac{1}{2}\frac{\overleftarrow{\delta}S_c}{\delta \chi_+(-k,t)}+\bar\eta(k,t),
\end{equation}
where $V(k)$ is a kernel used for convenience. In this case, we use the simple one
\begin{equation}
\label{345}
V(k)=-k_{\nu}\gamma^{\hat \nu}.
\end{equation}
Due to the kernel, the partion function of the Stochastic quantization change slightly 
\begin{equation}
\label{344}
Z=\int \mathcal D\eta \mathcal D\bar\eta \exp\left( -\int dtd^dk \delta^{\alpha\beta} \bar\eta_\alpha(k,t^\prime)V^{-1}(k)\eta_\beta(-k,t^\prime)\right).
\end{equation}

For explicit caculation, we put the classcial action and the kernel into the Langevin equations, then the Langevin equation becomes
\begin{equation}
\label{346}
\frac{\partial \chi_+(k,t)}{\partial t}=\partial_t\log\left(t^{\frac{1}{2}}K_{m-\frac{1}{2}}(|k|t)\right)\chi_+(k,t)+\eta(k,t),
\end{equation}

\begin{equation}
\label{347}
\frac{\partial \bar\chi_+(k,t)}{\partial t}=\bar\chi_+(k,t)\partial_t\log\left(t^{\frac{1}{2}}K_{m-\frac{1}{2}}(|k|t)\right)+\bar\eta(k,t).
\end{equation}

The solutions of the Langevin equations are as follows,

\begin{equation}
\label{348}
\chi_+(k,t)=\int^t\frac{t^{\frac{1}{2}}K_{m-\frac{1}{2}}(|k|t)}{t^{\prime\frac{1}{2}}K_{m-\frac{1}{2}}(|k|t^\prime)}\eta(k,t^\prime)dt^\prime,
\end{equation}

\begin{equation}
\label{349}
\bar\chi_+(k,t)=\int^t\frac{t^{\frac{1}{2}}K_{m-\frac{1}{2}}(|k|t)}{t^{\prime\frac{1}{2}}K_{m-\frac{1}{2}}(|k|t^\prime)}\bar\eta(k,t^\prime) dt^\prime.
\end{equation}

With the solutions of the $\chi$ fields, we can compute the two-point correlation of the $\chi_+$ fields in the stochastic quantizaiton side,

\begin{eqnarray}
\label{350}
\langle \chi_{\alpha,+}(k,t)\bar\chi_{\beta,+}(k^\prime,t)\rangle&=&\int^t_{t_0}\int^{t}_{t_0}\frac{t K_{m-\frac{1}{2}}(|k|t)K_{m-\frac{1}{2}}(|k|^\prime t)}{\tilde t^{\frac{1}{2}}K_{m-\frac{1}{2}}(|k|\tilde t^)\tilde t^{\prime\frac{1}{2}}K_{m-\frac{1}{2}}(|k|\tilde t^\prime)}\langle \eta_{k,\alpha}(\tilde t)\bar\eta_{k^\prime,\beta}(\tilde t^\prime) \rangle d\tilde t d\tilde t^\prime \\ \nonumber
&=&-k_\mu \gamma^{\hat \mu}_{\alpha\beta}\delta(k+k^\prime)tK_{m-\frac{1}{2}}(|k|t)^2\int^t_{t_0}d\tilde t \frac{1}{tK_{m-\frac{1}{2}}(|k|\tilde t)^2}\\ \nonumber
&=&-k_\mu \gamma^{\hat \mu}_{\alpha\beta}\delta(k+k^\prime)t\left( K_{m-\frac{1}{2}}(|k|t) I_{m-\frac{1}{2}}(|k|t)+ K_{m-\frac{1}{2}}(|k|t)^2\frac{ I_{m-\frac{1}{2}}(|k|t_0)}{ K_{m-\frac{1}{2}}(|k|t_0)} \right) \\ \nonumber
&=&-E_\alpha(k) \delta_{\alpha\beta}\delta(k+k^\prime)t K_{m-\frac{1}{2}}(|k|t) I_{m-\frac{1}{2}}(|k|t)\left( 1+\frac{ K_{m-\frac{1}{2}}(|k|t)}{\tilde\Delta_\alpha I_{m-\frac{1}{2}}(|k|t)} \right).
\end{eqnarray}
Now we already get the two-point correlation of both theories. However, we should set the initial time $t_0$ of stochastic quantizaiton to match the results of two theories. In this case, the initial time $t_0=\frac{1}{|k|}\left( \frac{K_{m-\frac{1}{2}}}{I_{m-\frac{1}{2}}}\right) ^{-1} (-\tilde\Delta_\alpha)$ is used.
We can check out the relation between stochastic quantization and AdS/CFT,
\begin{equation}
\label{351}
\langle \chi_{\alpha,+}(k,t) \bar \chi_{\beta,+}(k,t) \rangle_H^{-1} = \langle \chi_{\alpha,+}(k,t) \bar \chi_{\beta,+}(k,t) \rangle_S^{-1}-\frac{1}{2} \frac{\overrightarrow\delta}{\delta \bar \chi_{\alpha,+}(-k,t)} 
S_c \frac{\overleftarrow\delta}{\delta \chi_{\beta,+}(-k,t)}.
\end{equation}
Although the calculation seems boring, but the result is simple. The Langevin approach is consistent with the result of the AdS/CFT.

\subsection{Fokker-Plank approach}
 The second way which can be used to compute the double trace operator is  the Fokker-Plank way. The general form of the Fokker-Plank Lagrangian is gotten from the stochastic partition function of fermions,
\begin{eqnarray}
\label{352}
\mathcal L_{FP}&= &
\frac{\partial \bar \chi_\alpha(k,t)}{\partial t} 
[V^{-1}]^{\alpha\beta}(k)\frac{\partial  \chi_\beta(-k,t)}{\partial t}
\\ \nonumber
&-&\frac{1}{2} \bar \chi_\alpha(k,t)\partial_t ^2\log\left(t^{\frac{1}{2}}K_{m-\frac{1}{2}}(|k|t)\right) \chi_\beta(-k,t)+ \frac{1}{4}\frac{\overleftarrow\delta S_c}{\delta \chi_\alpha}V_{\alpha\beta}(k)\frac{\overrightarrow\delta S_c}{\delta \bar\chi_\beta}.
\end{eqnarray}
where we use the same kernel $V(k)$ with the Langevin's approach. As a result of putting the kernel and the classical action into the Fokker-Plank Lagrangian, the Fokker-plank action is given by
\begin{eqnarray}
\label{353}
S_{FP} &=& -\int dtd^dk\left[ \partial_t \bar\chi_+(k,t)\frac{k_{\hat \mu}\gamma^{\hat \mu}}{|k|^2}\partial_t\chi_+(-k,t)+\bar\chi_+(k,t)k_{\hat \mu}\gamma^{\hat \mu}\chi_+(-k,t)\right. \\ \nonumber
&+& \left. \frac{m(m-1)}{t^2}\bar \chi_+(k,t)\frac{k_{\hat \mu}\gamma^{\hat \mu}}{|k|^2}\chi_+(-k,t)\right].
\end{eqnarray}
You can check that the Fokker-Plank action is same with the newly defined bulk action in the AdS/CFT section (\ref{225}) if we put the suggested condition (1) $r=t$ referred in the introduction. From the Fokker-Plank action, we could compute equations of motion of the $\chi_+$ and $\bar\chi_+$ fields by applying the principle of least action. The result for the $\chi_+$ field is
\begin{equation}
\label{354}
\partial^2_t \chi_\alpha(k,t)-\left( k^2+\frac{m(m-1)}{t^2} \right) \chi_\alpha(k,t)=0.
\end{equation}
where $\alpha$ are spin indeces.
The solutions of the equations of motion also are given as the linear combination of the modified bessel functions. The result of inserting the equations of motion into the Fokker-Plank action is the following boundary terms, 
\begin{equation}
\label{355}
S_{FP}=-\int d^dk \left(\left. \bar \chi(k,t^\prime)\frac{k_\mu \gamma^{\hat \mu}}{k^2}\frac{\partial \chi(-k,t^\prime)}{\partial t^\prime} \right)\right|^{t^\prime=t}_{t^\prime=t_0}.
\end{equation}

For convenience, we take out the dependancy of $t'$ from the $\chi_+$ field,
\begin{equation}
\label{356}
\chi_{\alpha,+}(k,t^\prime)=\zeta^{-1}_{\alpha,+}(k,t)\zeta_{\alpha,+}(k,t^\prime)\chi_{\alpha,+}(k,t),
\end{equation}

\begin{equation}
{\rm where \ \ }
\label{357}
\zeta_{\alpha,+}(k,t)=\tilde N_\alpha[K_{m-\frac{1}{2}}(|k|t)+\tilde\Delta_\alpha(k)I_{m-\frac{1}{2}}(|k|t)]|\alpha\rangle.
\end{equation}

And likewise the $\bar\chi_+$ field also can be written as
\begin{equation}
\label{358}
\bar\chi_{\alpha,+}(k,t^\prime)=\bar\chi_{\alpha,+}(k,t)\bar\zeta_{\alpha,+}^{-1}(k,t)\bar\zeta_{\alpha,+}(k,t^\prime),
\end{equation}

\begin{equation}
{\rm where \ \ }
\label{359}
\bar\zeta_{\alpha,+}\alpha(k,t)=\bar\zeta_{\alpha,0}(k)K_{m-\frac{1}{2}}(|k|t)+\bar\zeta_{\alpha,1}(k)I_{m-\frac{1}{2}}(|k|t)(|k|t),
\end{equation}

Finally we put the $\chi_+$ and $\bar \chi_+$ fields into the Fokker-Plank action,
\begin{eqnarray}
\label{360}
S_{FP}&=&\sum_{\alpha,\beta}\int d^dk\left[ \bar\chi_{\alpha,+}(k,t)\left(\frac{-[k_\mu\gamma^{\hat\mu}]_{\alpha\beta}}{|k|^2}\right)\zeta_{\beta,+}^{-1}(-k,t)\frac{\partial \zeta_{\beta,+}(-k,t)}{\partial t}\chi_{\beta,+}(-k,t) \right], \\ \nonumber
&=&\sum_{\alpha,\beta}\int d^dk\left[ \bar\chi_{\alpha,+}(k,t)\left(\frac{-E(k)\delta_{\alpha\beta}}{|k|^2}\zeta_{\beta,+}^{-1}(-k,t)\frac{\partial \zeta_{\beta,+}(-k,t)}{\partial t}\right)\chi_{\beta,+}(-k,t) \right], \\ \nonumber
&=&\sum_{\alpha,\beta}\int d^dk\left[ \bar\chi_{\alpha,+}(k,t)\left(\frac{E(k)\delta_{\alpha\beta}}{|k|^2}|k|\frac{K_{m+\frac{1}{2}}(|k|\epsilon)-\Delta_{\beta,+}I_{m+\frac{1}{2}}(|k|\epsilon)}{K_{m-\frac{1}{2}}(|k|\epsilon)
 +\Delta_{\beta,+}I_{m-\frac{1}{2}}(|k|\epsilon)} \right.\right. \\ \nonumber
&-&\left. \left.\frac{m}{r} \right)\chi_{\beta,+}(-k,t) \right], \\ \nonumber
&=&\int d^dk  \bar\chi_+(k,t)\mathcal D(k,t)\chi_+(-k,t),
\end{eqnarray}
where we impose the condition on $t_0$ which is same with the Langevin approach,

\begin{equation}
\label{361}
t_0=\frac{1}{|k|}\left( \frac{K_{m-\frac{1}{2}}}{I_{m-\frac{1}{2}}}\right) ^{-1} (-\tilde\Delta_\alpha)
\end{equation}
Here, the boundary terms vanish at the initial time $t=t_0$. As we expected, the $\chi^2$ coefficient $\mathcal D(k,t)$ of stochastic quantization is same with the $\chi^2$ coefficient of AdS/CFT(\ref{236}) if we change $r\rightarrow t$. The constant term and the one point function terms of the $\chi$ and $\bar\chi$ fields disappear because of the initial condtion, but if we set other initial time, then we can also check the other terms, in the trial function of the boundary action (\ref{228}).

\section*{Acknowledgement}
S.P.M would like to thank his adviser Jae-hyuk Oh for many useful discussions.
This work was supported by the National Research Foundation of Korea(NRF) grant funded by the Korea government(MSIP) (No. 2016R1C1B1010107).


\begin{thebibliography}{9}


\bibitem{1} 
  J.~H.~Oh and D.~P.~Jatkar,
  JHEP {\bf 1211}, 144 (2012)
  doi:10.1007/JHEP11(2012)144
  [arXiv:1209.2242 [hep-th]].

\bibitem{2} 
  J.~H.~Oh,
  Int.\ J.\ Mod.\ Phys.\ A {\bf 29}, 1450082 (2014)
  doi:10.1142/S0217751X14500821
  [arXiv:1310.0588 [hep-th]].

\bibitem{3} 
  D.~P.~Jatkar and J.~H.~Oh,
  JHEP {\bf 1310}, 170 (2013)
  doi:10.1007/JHEP10(2013)170
  [arXiv:1305.2008 [hep-th]].

\bibitem{4} 
  V.~Balasubramanian, P.~Kraus, A.~E.~Lawrence and S.~P.~Trivedi,
  Phys.\ Rev.\ D {\bf 59}, 104021 (1999)
  doi:10.1103/PhysRevD.59.104021
  [hep-th/9808017].

\bibitem{5} 
  I.~Heemskerk and J.~Polchinski,
  JHEP {\bf 1106}, 031 (2011)
  doi:10.1007/JHEP06(2011)031
  [arXiv:1010.1264 [hep-th]].

\bibitem{6} 
  T.~Faulkner, H.~Liu and M.~Rangamani,
  JHEP {\bf 1108}, 051 (2011)
  doi:10.1007/JHEP08(2011)051
  [arXiv:1010.4036 [hep-th]].

\bibitem{7} 
  D.~S.~Mansi, A.~Mauri and A.~C.~Petkou,
  Phys.\ Lett.\ B {\bf 685}, 215 (2010)
  doi:10.1016/j.physletb.2010.01.033
  [arXiv:0912.2105 [hep-th]].

\bibitem{8}
P.H. Damgaard and H. Huffel, Stochastic quantization. PHYSICS REPORTS (Review Section of Physics Letters) 152, Nos. 5 and 6 (1987) 

\bibitem{9} 
  E.~Witten,
  hep-th/0112258.

\bibitem{10} 
  D.~Elander, H.~Isono and G.~Mandal,
  JHEP {\bf 1111}, 155 (2011)
  doi:10.1007/JHEP11(2011)155
  [arXiv:1109.3366 [hep-th]].

\bibitem{11} 
  J.~N.~Laia and D.~Tong,
  JHEP {\bf 1111}, 131 (2011)
  doi:10.1007/JHEP11(2011)131
  [arXiv:1108.2216 [hep-th]].

\end{thebibliography}
\end{document}